\documentclass[a4paper,12pt]{article}
\pdfoutput=1
\usepackage{graphicx, rotating}
\usepackage{hyperref}
\usepackage{slashed}
\usepackage{ifpdf}

\ifx\pdfoutput\undefined
\usepackage[dvips,bookmarks=false]{hyperref}	
\else
\usepackage{hyperref}	
\fi
\hypersetup{colorlinks,bookmarksopen,bookmarksnumbered,citecolor=verdes,
linkcolor=blus,pdfstartview=FitH,urlcolor=rossos}
\def\hhref#1{\href{http://arxiv.org/abs/#1}{#1}} 

\newcommand{\cm}{\,{\rm cm}}

\usepackage{multicol}
\usepackage{color}
\definecolor{rosso}{cmyk}{0,1,1,0.4}
\definecolor{rossos}{cmyk}{0,1,1,0.55}
\definecolor{rossoc}{cmyk}{0,1,1,0.2}
\definecolor{blu}{cmyk}{1,1,0,0.3}
\definecolor{blus}{cmyk}{1,1,0,0.6}
\definecolor{bluc}{cmyk}{1,1,0,0.1}
\definecolor{verde}{cmyk}{0.92,0,0.59,0.25}
\definecolor{verdec}{cmyk}{0.92,0,0.59,0.15}
\definecolor{verdes}{cmyk}{0.92,0,0.59,0.4}

\font\tenrsfs=rsfs10 at 12pt
\font\sevenrsfs=rsfs7
\font\fiversfs=rsfs5
\newfam\rsfsfam
\textfont\rsfsfam=\tenrsfs
\scriptfont\rsfsfam=\sevenrsfs
\scriptscriptfont\rsfsfam=\fiversfs
\def\mathscr#1{{\fam\rsfsfam\relax#1}}

\newcommand{\fig}[1]{~\ref{fig:#1}}
\newcommand{\eq}[1]{~{\rm (\ref{eq:#1})}}

\newcommand{\GeV}{\,{\rm GeV}}
\newcommand{\TeV}{\,{\rm TeV}}

\def\circa#1{\,\raise.3ex\hbox{$#1$\kern-.75em\lower1ex\hbox{$\sim$}}\,}

\newcommand{\eqref}[1]{(\ref{#1})}

\newcommand{\NP}{Nucl. Phys.}

\newcommand{\beq}{\begin{equation}}
\newcommand{\eeq}{\end{equation}}

\def\circa#1{\,\raise.3ex\hbox{$#1$\kern-.75em\lower1ex\hbox{$\sim$}}\,}
\makeatletter

\def\art{\@ifnextchar[{\eart}{\oart}}
\def\eart[#1]#2#3#4#5#6{{\rm #2}, {#3 #4} {\rm (#6) #5} [arXiv:\-{\hhref{#1}}]}
\def\hepart[#1]#2{{\rm #2, arXiv:\-\hhref{#1}}}
\newcommand{\oart}[5]{{\rm #1}, {#2 #3} {\rm (#5) #4}}

\newcounter{alphaequation}[equation]
\def\thealphaequation{\theequation\hbox to
0.6em{\hfil\alph{alphaequation}\hfil}}
\def\eqnsystem#1{
\def\@eqnnum{{\rm (\thealphaequation)}}
\def\@@eqncr{\let\@tempa\relax \ifcase\@eqcnt \def\@tempa{& & &} \or
\def\@tempa{& &}\or \def\@tempa{&}\fi\@tempa
\if@eqnsw\@eqnnum\refstepcounter{alphaequation}\fi
\global\@eqnswtrue\global\@eqcnt=0\cr}
\refstepcounter{equation} \let\@currentlabel\theequation \def\@tempb{#1}
\ifx\@tempb\empty\else\label{#1}\fi
\refstepcounter{alphaequation}
\let\@currentlabel\thealphaequation
\global\@eqnswtrue\global\@eqcnt=0 \tabskip\@centering\let\\=\@eqncr
$$\halign to \displaywidth\bgroup \@eqnsel\hskip\@centering
$\displaystyle\tabskip\z@{##}$&\global\@eqcnt\@ne
\hskip2\arraycolsep\hfil${##}$\hfil& \global\@eqcnt\tw@\hskip2\arraycolsep
$\displaystyle\tabskip\z@{##}$\hfil
\tabskip\@centering&\llap{##}\tabskip\z@\cr}
\def\endeqnsystem{\@@eqncr\egroup$$\global\@ignoretrue} \makeatother

\begin{document}

\begin{center}
IFUP-TH/2009-25\hfill  CERN-PH-TH/2009-238

\bigskip\bigskip\bigskip

{\huge\bf\color{magenta}
Robust implications on Dark Matter\\[4mm]
from the first {\sc Fermi} sky $\gamma$ map}\\

\medskip
\bigskip\color{black}\vspace{0.6cm}
{
{\large\bf Michele Papucci$^a$ {\rm and} Alessandro Strumia$^{bc}$}
}
\\[7mm]
{\it $^a$ Institute for Advanced Study, Princeton, NJ 08540} \\
{\it $^b$ Dipartimento di Fisica dell'Universit{\`a} di Pisa and INFN, Italia} \\
{\it $^c$ CERN, PH-TH, CH-1211, Gen\`eve 23, Suisse}

\bigskip\bigskip\bigskip\bigskip

{
\centerline{\large\bf Abstract}

\begin{quote}
We derive robust model-independent bounds on Dark Matter (DM) annihilations and decays
from the first year of {\sc Fermi} $\gamma$-ray observations of the whole sky.
These bounds only have a mild dependence on the DM density profile
and allow the following DM interpretations of the PAMELA and {\sc Fermi} $e^\pm$ excesses:  
primary channels
$\mu^+\mu^-$, $ \mu^+\mu^-\mu^+\mu^-$ or $e^+e^-e^+e^-$.
An  isothermal-like density profile is needed for annihilating DM.
In all such cases, {\sc Fermi} $\gamma$ spectra must contain a significant DM component,
that may be probed in the future.\end{quote}}

\end{center}

\tableofcontents\newpage

\section{Introduction}
Recently the {\sc Fermi} collaboration released the first sky map of $\gamma$-rays up to energies below a few hundred $\GeV$~\cite{FERMIweb, FERMIgamma, FERMIdiffuse}.
From a particle physics point of view its main interest resides in the possible presence of a DM signal over the astrophysical background.
Excitingly enough, although no clear excess is present, a few theorists claim possible hints~\cite{Weiner}.
In this paper we do not address these issues, that will need a full understanding of the data 
(released photon data still contain a non-negligible contamination from mis-identified hadrons at energies around and above 100 GeV),
a proper modeling of the astrophysical backgrounds,
subtraction of identified point-sources,
and maybe more statistics.

Here we take a different approach: using the available data we derive robust bounds on DM annihilations and decays,
by only demanding the DM-induced $\gamma$-ray flux to be below the observed flux.
Since DM is neutral, $\gamma$'s are only produced at higher order in the QED coupling from various processes:
we only include those contributions that can be computed in a model-independent way.

We do not attempt to subtract the astrophysical and instrumental backgrounds, nor point sources: many of these subtractions can be performed only by assuming some astrophysical model, reducing the robustness of the bounds. Whenever possible, we will also comment on and try to quantify the residual uncertainties of our results.
Progress on understanding the backgrounds  can only  render  our bounds more stringent, presumably by order one factors.
A Dark Matter signal could be lurking just below our bounds.

\medskip

An interesting application of our results is checking whether the DM  interpretations of the $e^\pm$ excesses
observed by PAMELA~\cite{PAMELA}, {\sc Fermi}~\cite{FERMI}  and ATIC~\cite{ATIC-2}
give an Inverse Compton (IC) photon flux compatible with $\gamma$-ray observations.
Indeed, according to DM interpretations, the $e^\pm$ excess should be present
everywhere in the DM halo (rather than only locally, as if the  $e^\pm$ excess is due to a nearby astrophysical source
such as a pulsar),
giving rise to an unavoidable associated $\gamma$-ray signal: $e^\pm$ produced by DM
loose essentially all their energy by Compton up-scattering ambient light,
giving rise to a photon flux at the level of {\sc Fermi} sensitivity.
Being a diffuse signal, one can consider regions of the sky that have smaller astrophysical uncertainties.
By doing so, we find that many DM interpretations of the $e^\pm$ excesses are already excluded by this bound
(future improvements will not change this conclusion); moreover,
in some cases the expected effect is at the level of the observed flux
(future improvements should allow to test it).

\bigskip

The plan of the paper is the following: in Section 2 we present technical details of our analysis; in Section 3 we present results for DM annihilations 
and in Section 4 for DM decays. Results are summarized in the Conclusions.

\section{Technical introduction}

We consider DM annihilations (parameterized by the DM DM cross section $\sigma v$)
or decays (parameterized by the DM decay rate $\Gamma=1/\tau$)
into the following
set of primary DM particles:
\beq 2e, 2\mu,2\tau, 4e,4\mu,4\tau, 2q, 2b, 2t, 2h,2W\eeq
where $2e$ stands for $e^+e^-$ and $4e$ stands for $VV$, where $V$ is an hypotethical
new light particle, whose mass  we take to be $m\sim1\GeV$, that decays into SM final states, as e.g.\ in the model of~\cite{Nima}.
We denote as $M$ the DM mass. For the Higgs boson, we assume a mass $m_h=115\GeV$.
Concerning all the details not specified here, we follow~\cite{MPSV}.

\subsection{Astrophysics}\label{astro}
We consider  the following  Milky Way DM tentative density profiles $\rho(r)$~\cite{rho(r)}: 
\beq \frac{\rho(r)}{\rho_\odot} = \left\{\begin{array}{ll}
(1+r_\odot^2/r_s^2)/(1+r^2/r_s^2) & \hbox{isothermal, $r_s = 5\,{\rm kpc}$}\\
(r_\odot/r)(1+r_\odot/r_s)^2/(1+r/r_s)^2 & \hbox{NFW, $r_s = 20\,{\rm kpc}$}\\
\exp(-2[(r/r_s)^\alpha - (r_\odot/r_s)^\alpha]/\alpha)& \hbox{Einasto, $r_s = 20\,{\rm kpc}$, $\alpha=0.17,$}\\
\end{array}
\right.
\eeq
keeping fixed the local DM density $\rho(r=r_\odot\approx 8.5\,{\rm kpc}) = \rho_\odot=0.3\GeV/\cm^3$.
NFW and Einasto profiles are favored by $N$-body simulations, isothermal-like profiles
by observations of spiral galaxies~\cite{Salucci}.

When assessing the residual uncertainties of our bounds, we will also consider the possibility of a disk-like component for the DM~\cite{ddisk}. For this ``Dark Disk'' we will assume a profile $\rho_{\rm disk}(r,z) \propto \exp(r/r_d)\sinh(z/z_d)^2$ and we will vary $r_d$ between 5 and 15 kpc, $z_d$ between 0.2 and 10 kpc, and the fraction of DM in the Dark Disk at solar position from 0 to 50\%.

\medskip

Regarding the diffusion of $e^\pm$ in the Milky Way, we consider the min, med, max propagation models of~\cite{minmedmax}
characterized by the following astrophysical parameters:
\beq \begin{tabular}{ccccc}
Model  & $\delta$ & $K_0$ in kpc$^2$/Myr & $L$ in kpc \\
\hline 
min  & 0.85 &  0.006 & 1 \\
med  & 0.70 &  0.011 & 4   \\
max  & 0.46 &  0.076 & 15
\end{tabular}\label{eq:proparam}\quad . \eeq
The diffusion coefficient $K = K_0 E^\delta$ is assumed to be constant inside a
cylinder with height $2L$ centered on the galactic plane and radius 20 kpc, 
and infinitely large outside.

When assessing the dependence of our bounds on the size of the diffusion zone, we will consider also the more realistic case in which the diffusion coefficient depends on the distance from the galactic plane~\cite{dario}, $K(E,z)=K_0 E^\delta \exp (|z|/z_h)$. where $z_h$ effectively determines the thickness of the diffusion zone, that dies off gradually instead of turning off abruptly at $|z|=L$.

 \begin{figure}[t]
\begin{center}
$$\includegraphics[width=\textwidth]{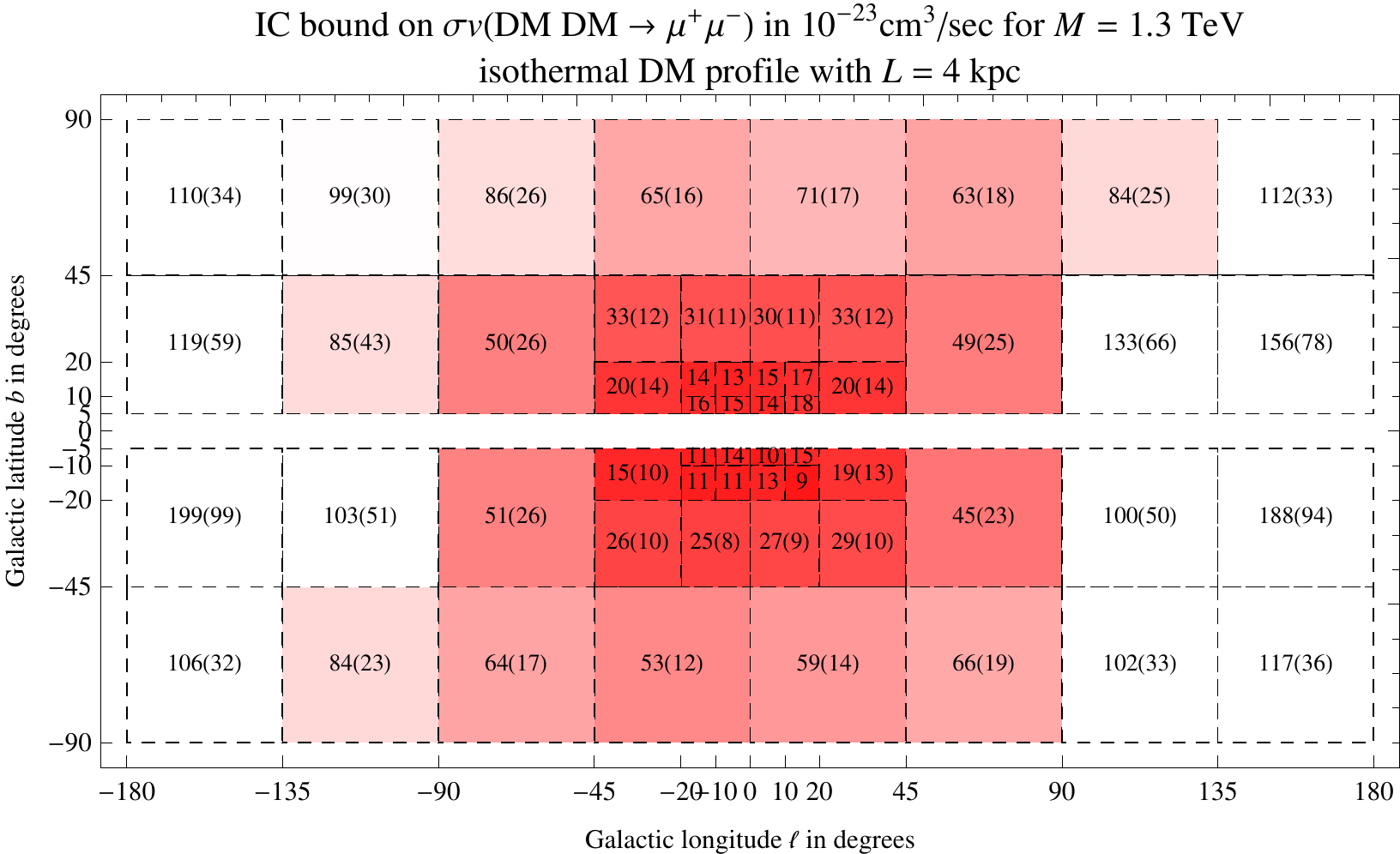}$$
\caption{\em Subdivision of the sky and example of a bound from each different region.
In parenthesis, the same bound is computed neglecting both the diffusion of $e^\pm$ and the finite volume of the Milky Way
diffusion halo. 
}
\label{fig:mosaic}
\end{center}
\end{figure}

\subsection{Fitting the {\sc Fermi} sky map}
We divide the {\sc Fermi} $\gamma$-ray sky, parameterized by galactic longitude $\ell$ and latitude $b$
($\ell=b=0$ corresponds to the Galactic Center, GC) into several regions, depicted in Fig.\fig{mosaic}.
We extract the photon spectrum within each region from the first public release of the {\sc Fermi} $\gamma$-ray
data~\cite{FERMIweb}. The details of the extraction are summarized in the Appendix.
The {\sc Fermi} collaboration published the energy spectra below 100~GeV in a few regions~\cite{FERMIgamma};
we checked that our procedure reproduces the {\sc Fermi} results at low and intermediate energies.
We do not subtract point sources, but we exclude the region most polluted by astrophysical sources,
the galactic plane, by restricting ourselves to $|b|>5^\circ$. Since the signal we are seeking to bound is not uniformly distributed on the sky, we consider a finer grid around the GC, where DM (but also astrophysical sources) concentrate. Conversely we use regions with increasingly larger areas towards the two poles, where the signal is expected to be smaller. Furthermore, we separately consider north-west, south-west, north-east and south-east regions, because the one less polluted by astrophysical sources will offer the best sensitivity.

\begin{figure}
\begin{center}
$$\includegraphics[width=0.5\textwidth]{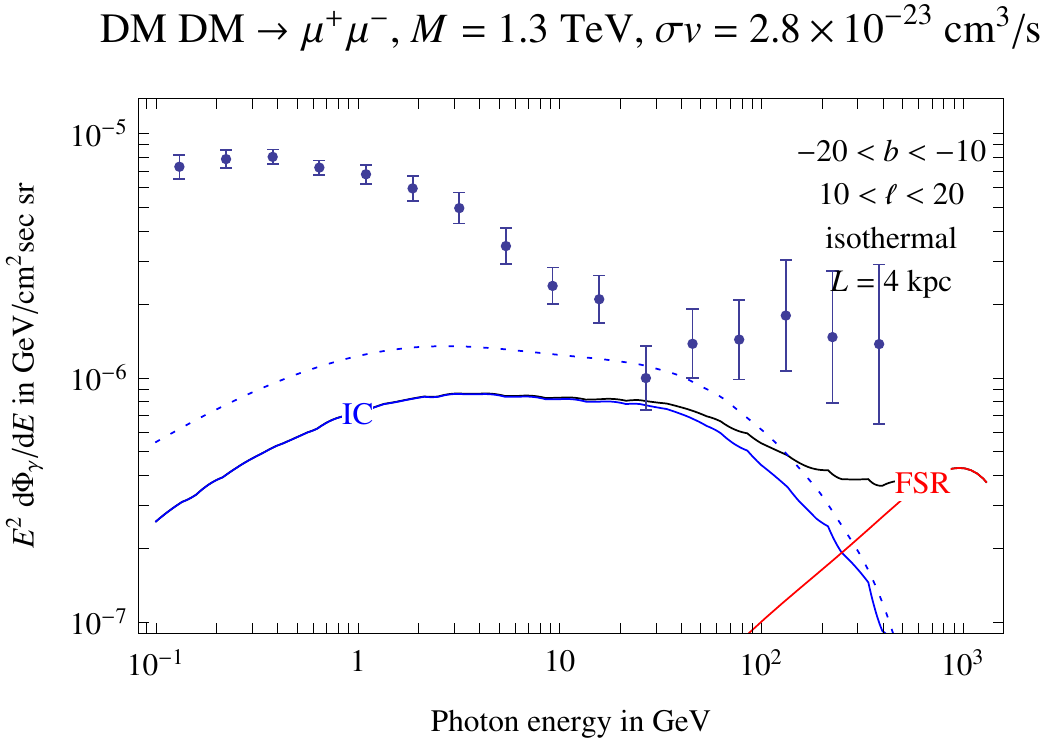}\quad
\includegraphics[width=0.485\textwidth]{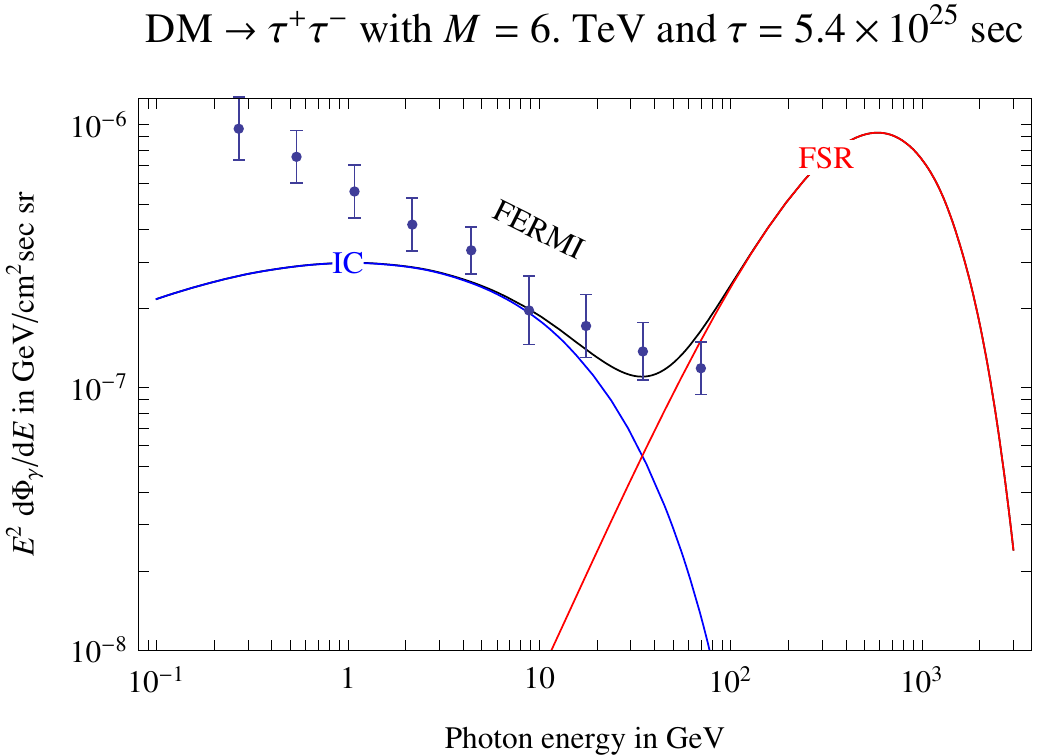}$$
\caption{\em Left: {\sc Fermi} data compared with an example of best-fit DM annihilation signal.
Photons above $\approx100 \GeV$ can be still contaminated by hadrons.
The dotted line shows Inverse Compton computed neglecting $e^\pm$ diffusion and the finite volume of the diffusion halo.
Right: {\sc Fermi} preliminary extra-galactic data~\cite{FERMIdiffuse} compared with an example of best-fit  DM decay signal.}
\label{fig:cosmosample}
\end{center}
\end{figure}

As the DM density profile is unknown, we do not know which region is most sensitive to DM.
A robust constraint is obtained by demanding that the minimal computable
DM-$\gamma$ spectrum in each region, $\Phi_i^{\rm DM} $ in energy bin $i$,
does not  exceed the measured $\gamma$ flux $ \Phi_i^{\rm exp}$ at $3\sigma$, for any energy bin
 and any region. Fig.\fig{mosaic} shows an example of the bounds on the DM annihilation cross sections
for each region.  
Fig.\fig{cosmosample}a compares the {\sc Fermi} data in the single region that gives the stronger bound
for this particular model  with the model prediction at its bets-fit point for the PAMELA and {\sc Fermi} $e^\pm$ excesses.

Various regions give comparable bounds. 
Thereby one can do slightly better, still maintaining the absolute robustness of the bounds, by combining all regions
in a global fit.
We impose the $3\sigma$ bound, $\chi^2 < 9$, where
\beq \chi^2 =\min_e \sum_i \frac{ (\Phi_i^{\rm DM}(E_i (1+e))
 - \Phi_i^{\rm exp})^2  }{\delta \Phi^2} \Theta(\Phi_i^{\rm DM} - \Phi_i^{\rm exp})
+\frac{e^2}{\delta e^2} \ ,\eeq
where $\Theta(x)=1$ if $x>0$ and 0 otherwise;
the sum runs over all angular and energy bins (with mean energy $E_i$);
the $\chi^2$ must be marginalized (minimized in Gaussian approximation) over  the energy-scale free parameter $e$;
the last term in the  $\chi^2$ accounts for the $\delta e \approx15\%$ {\sc Fermi} uncertainty on the energy scale;
that only has a minor effect on the bounds.

Typically, such global bound is a factor of few stronger than the bound obtained demanding
that no single point is exceeded at more than $3\sigma$.  Furthermore, there are conceptual advantages.
The `single point' bound depends on how we choose the energy and angular binning
and can be fully dominated by the single bin where a downward statistical fluctuation happened in the total rate.
On the other hand, the global fit does not depend on the binning (in the limit where it is dense enough);
e.g.\ one can even split one bin into two coincident bins without affecting the global fit.

\subsection{Computing $\gamma$'s from DM}
Since DM has no electric charge, $\gamma$-ray production from DM annihilations or decays occurs at higher order in the electromagnetic
coupling from many different processes with comparable rates:
i) bremsstrahlung   from charged particles and $\pi^0$ decays; ii) virtual emission, iii) loop effects;
iv) astrophysical processes involving other particles produced by DM.
We only consider the following two sources of $\gamma$-rays that can be robustly computed
in a model-independent way:
\begin{enumerate}
\item FSR$\gamma$, i.e.\ {\em Final State Radiation} emitted by the primary DM annihilation or in subsequent decays. With a slight abuse of terminology, we will include in this contribution also the photons from hadronic decays.
This gives photons with the largest $E_\gamma \sim M$.

\end{enumerate}
Primary channels such as $2\tau, 2q, 2W$ give rise to $\pi^0$s that decay as
$\pi^0\to 2\gamma$ giving a $\gamma$ yield larger than channels (such as $2e$ or $2\mu$)
that only produce $\gamma$ from bremsstrahlung.
Models involving new neutral light particles give the smallest $\gamma$ yield~\cite{BCST,MPV}.
We do not consider electroweak bremsstrahlung, that cannot be computed
in a model-independent way: one needs to know in which electroweak multiplets the DM lies.
The large corrections found in~\cite{Serpico} if $M\gg 4\pi v$
are only present for those decay or annihilation channels that arise thanks to a non-vanishing Higgs vev $v$.
All the channels we consider can be realized, in appropriate DM models, 
as effective operators that do not involve the Higgs
(e.g.\ as effective operators where a DM pair couples to a vector leptonic current, or to the $W,Z$ field-strength squared),
such that electroweak bremsstrahlung remains one higher order effect, as in~\cite{Edjso}.

\medskip

At lower photon energies, the DM$\gamma$ flux is dominated by the second source
of $\gamma$-rays that we consider~\cite{MPSV,IC}
\begin{enumerate}

\item[2.] IC$\gamma$, i.e.\ photons from Inverse Compon. DM gives rise to $e^\pm$ that loose most of their energy by
up-scattering galactic ambient light
(CMB  and starlight, partially rescattered by dust):
this {\em Inverse Compton} $e^\pm \gamma\to e^{\prime\pm} \gamma'$  process gives rise to $\gamma'$s with energy
$E_{\gamma'} \sim E_\gamma (E_e/m_e)^2\sim 30\GeV$.
\end{enumerate}
The IC energy loss process competes with energy losses due to synchrotron radiation in the galactic magnetic fields.
The rates of these two processes are respectively proportional to $u_\gamma (\vec x)$ and to $u_B (\vec x)= B^2/2$,
the energy densities in photons and in magnetic fields.  We take the values from~\cite{isrf,galprop}, and in particular we assume
\begin{equation}
\label{eq:Brz}
B(r,z) \approx 11\mu{\rm G} \cdot\exp(-r/10\,{\rm kpc} - |z|/2\,{\rm kpc}).
 \end{equation}
In this case $u_\gamma \gg u_B$ everywhere:
Inverse Compton is dominant, and it can be reliably computed as essentially all the $e^\pm$ energy goes into IC,
irrespectively of the precise galactic maps of $u_\gamma$ and of $u_B$.

Only large deviations from the maps we adopted can affect our bounds, so we comment about this possibility.
Both galactic radiation and magnetic fields are better known for our neighborhood than for other regions of the Galaxy, with the magnetic field being the most uncertain.  The most realistic worry is that magnetic fields
in the Inner Galaxy might be intense enough that one there has  $u_B \sim u_\gamma$,
weakening our bounds.   We checked the dependence on the magnetic field uncertainties by varying the scales 
in Eq.\eq{Brz} on which the galactic magnetic field changes both radially and vertically, while keeping its value at solar position fixed. We find that a factor of 2 variation in these scales changes the IC$\gamma$ fluxes at high latitudes by less than $10\div 20\%$ and in regions closer to the GC ($|b|<20^\circ$) by 60\%. Since these regions are relevant in our global fit, our bounds can be relaxed at most by a factor of $\sim1.5\div2$.

\medskip

In order to compute IC$\gamma$,
following~\cite{MPSV} we take into account the diffusion of $e^\pm$, with characteristic diffusion length of about $\lambda\sim 1\,{\rm kpc}$,
such that photons from IC are smeared in a region with angular size $\lambda/r_\odot \sim 10^\circ$:
even if  DM annihilations are  concentrated close to the GC, IC$\gamma$ are not.
Thereby {\sc Fermi} $\gamma$-ray data are more relevant than HESS data
both for the energy range and for the angular range they observe.

 \begin{figure}[p]
\begin{center}
$$\includegraphics[width=\textwidth]{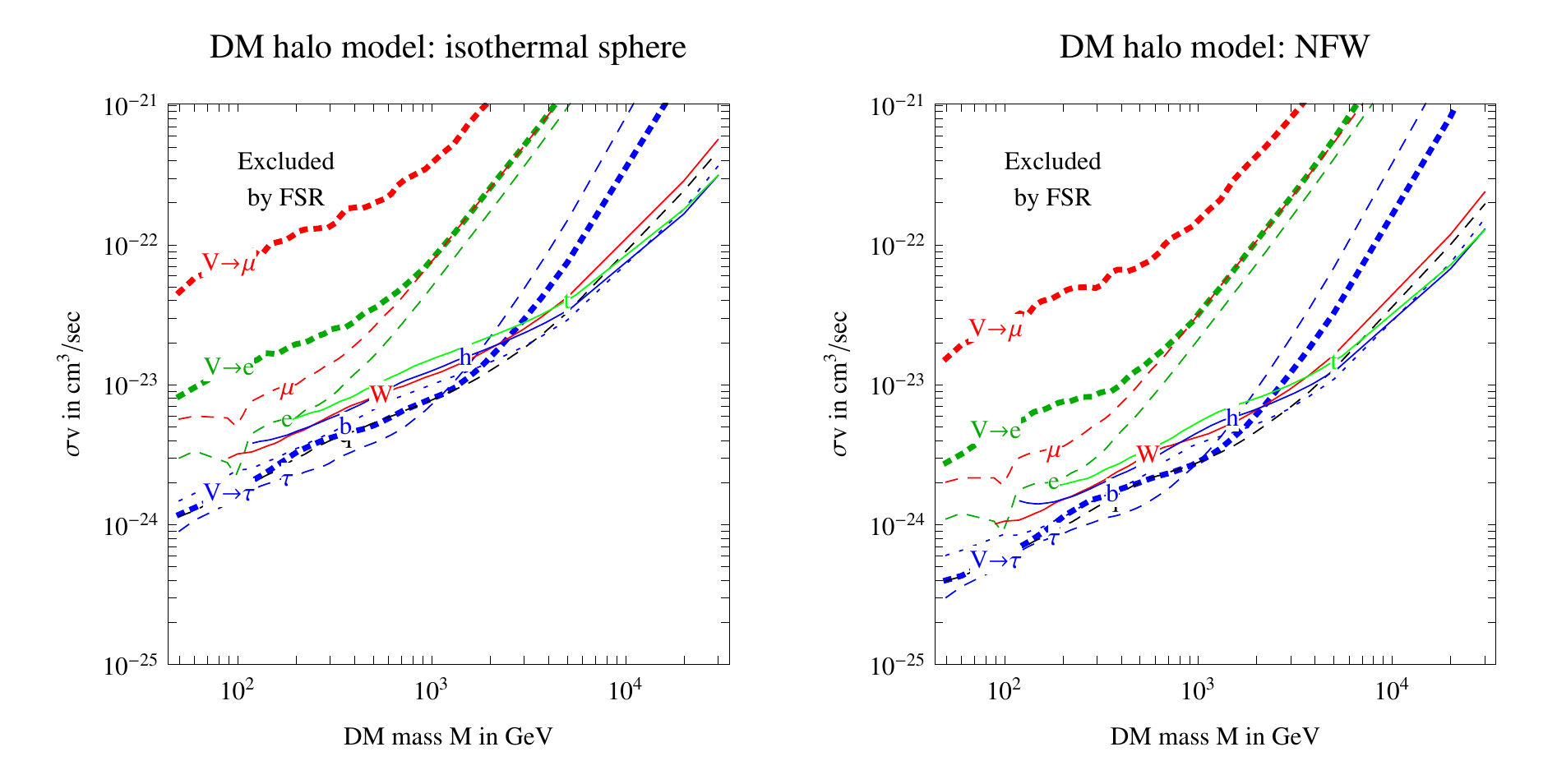}$$
\caption{\em {\sc Fermi} full-sky bounds on Final State Radiation $\gamma$-rays, for the DM annihilation modes
indicated along the curves.}
\label{fig:FSR}
\end{center}
\end{figure}

 \begin{figure}[p]
\begin{center}
$$\includegraphics[width=\textwidth]{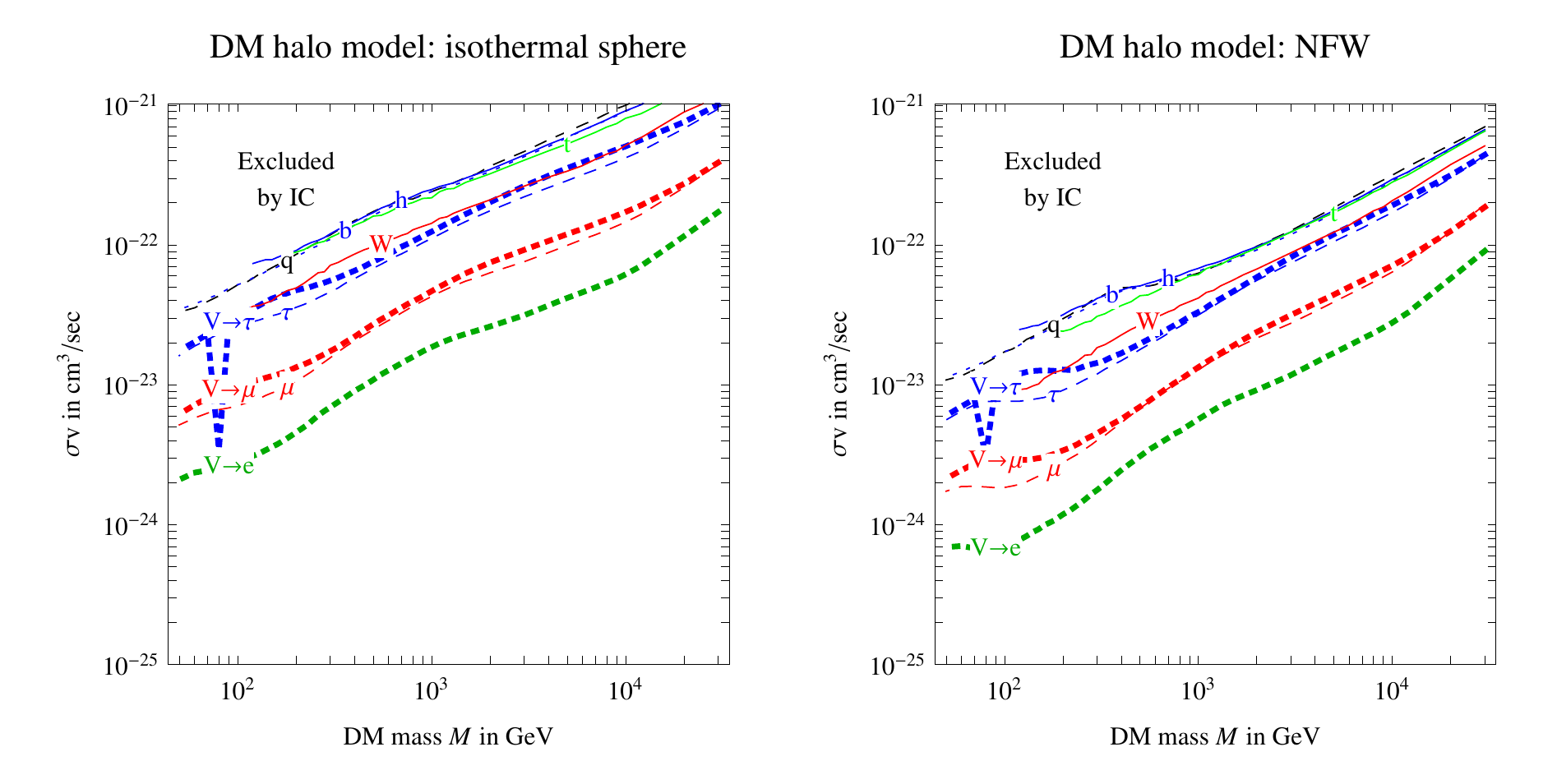}$$
\caption{\em {\sc Fermi} full-sky bounds on Inverse Compton $\gamma$-rays, for the DM annihilation modes
indicated along the curves.}
\label{fig:IC}
\end{center}
\end{figure}

 \begin{figure}
\begin{center}
$$\includegraphics[width=0.5\textwidth]{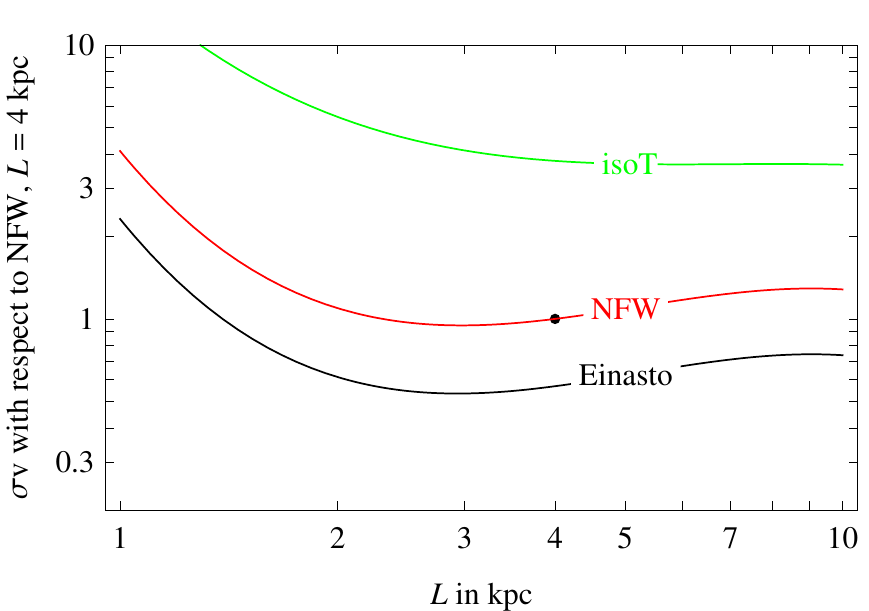}$$
\caption{\em Example of how the {\sc Fermi} IC$\gamma$ bound on $\sigma v$ changes as function
of the height $L$ of the diffusion volume for different DM profiles.
We here assumed DM annihilations into $\mu^+\mu^-$ with $M=1.3\TeV$, but this plot
would be almost the same for other DM models.}
\label{fig:ICL}
\end{center}
\end{figure}

 \begin{figure}[t]
\begin{center}
$$\includegraphics[width=\textwidth]{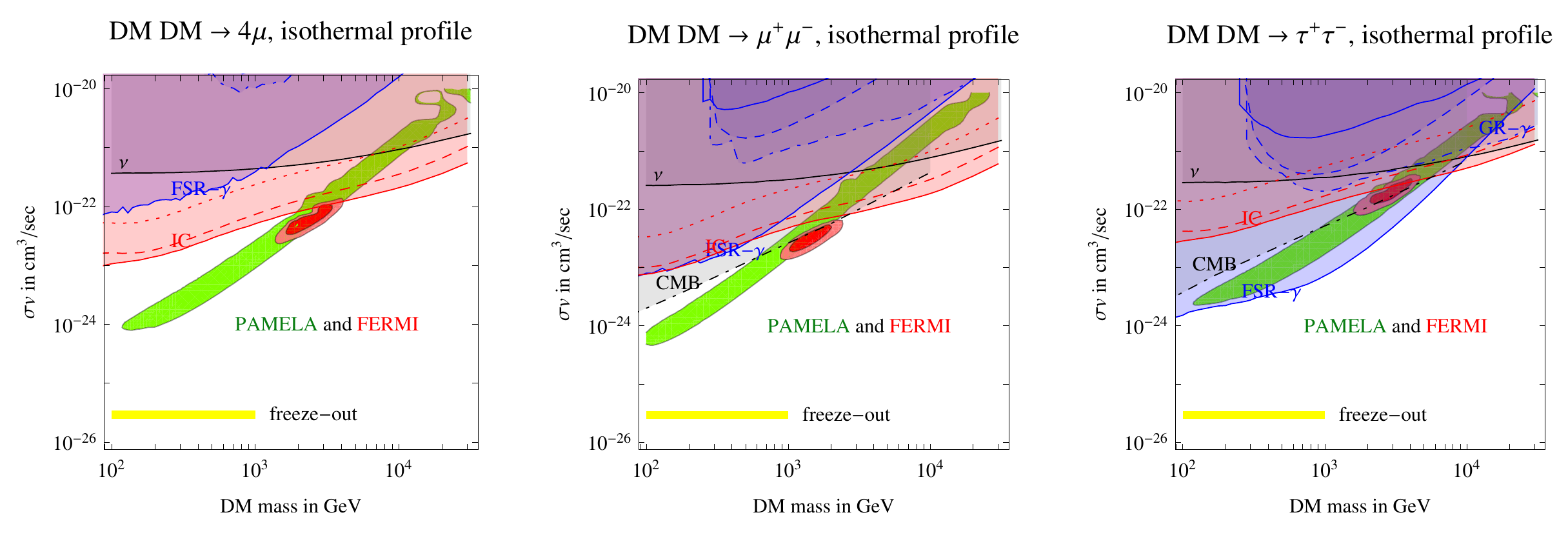}$$\vspace{-1cm}
$$\includegraphics[width=\textwidth]{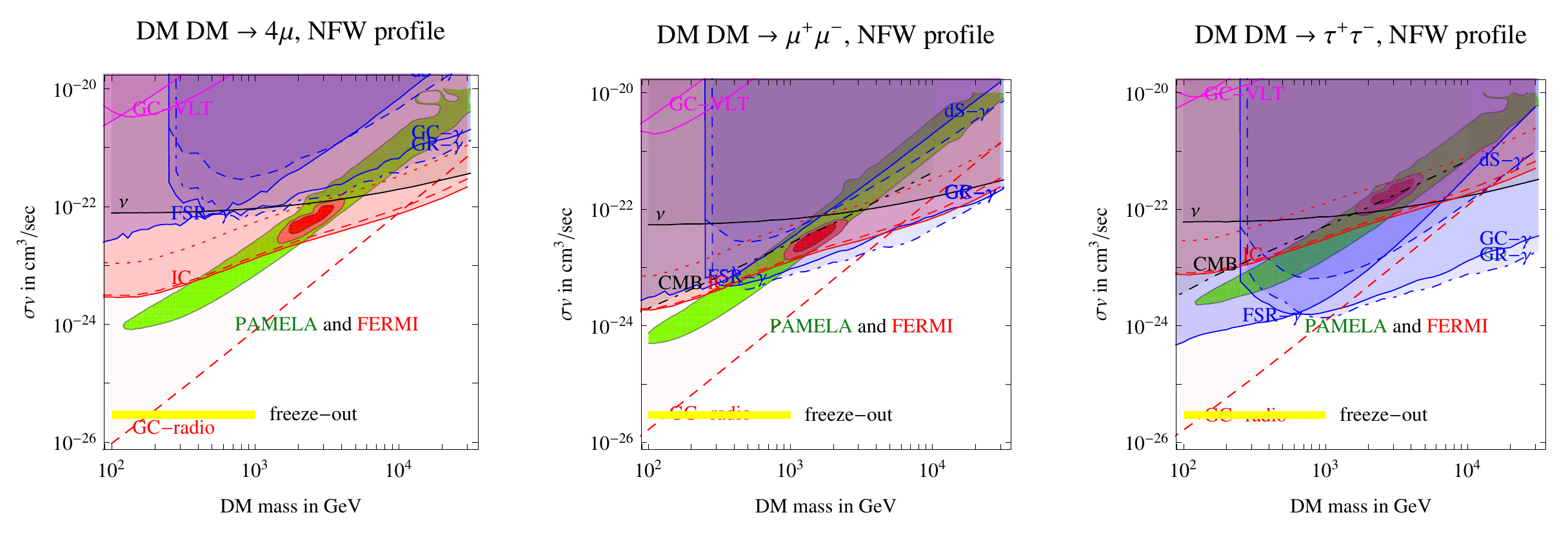}$$\vspace{-1cm}
\caption{\em {\bf Bounds on DM  annihilations into leptonic channels}. The {\sc Fermi} bounds are denoted as FSR$\gamma$ 
(continuous blue line)
and
IC$\gamma$ (red curves, for $L=1,2,4\,{\rm kpc}$ from upper to lower).
Other bounds are described in the text;
their labels appear along the corresponding lines only when these bounds are significant
enough to appear within the plots.
Cosmological freeze-out predicts $\sigma v\approx 3~10^{-26}\cm^3/\sec$ (lower horizontal band)
and connections with the hierarchy problem suggest $M\sim (10\div1000)\GeV$.
The region that can fit the $e^\pm$ excesses
survives only if DM annihilates into $e$'s or $\mu$'s and
DM has an isothermal profile.
All bounds are at $3\sigma$; 
the green bands are favored by PAMELA (at $3\sigma$ for 1 dof) and
the red ellipses by PAMELA, FERMI and HESS (at $3$ and $5\sigma$, 2 dof, as in~\cite{MPSV}).
}
\label{fig:ann}
\end{center}
\end{figure}

 \begin{figure}[t]
\begin{center}
$$\includegraphics[width=\textwidth]{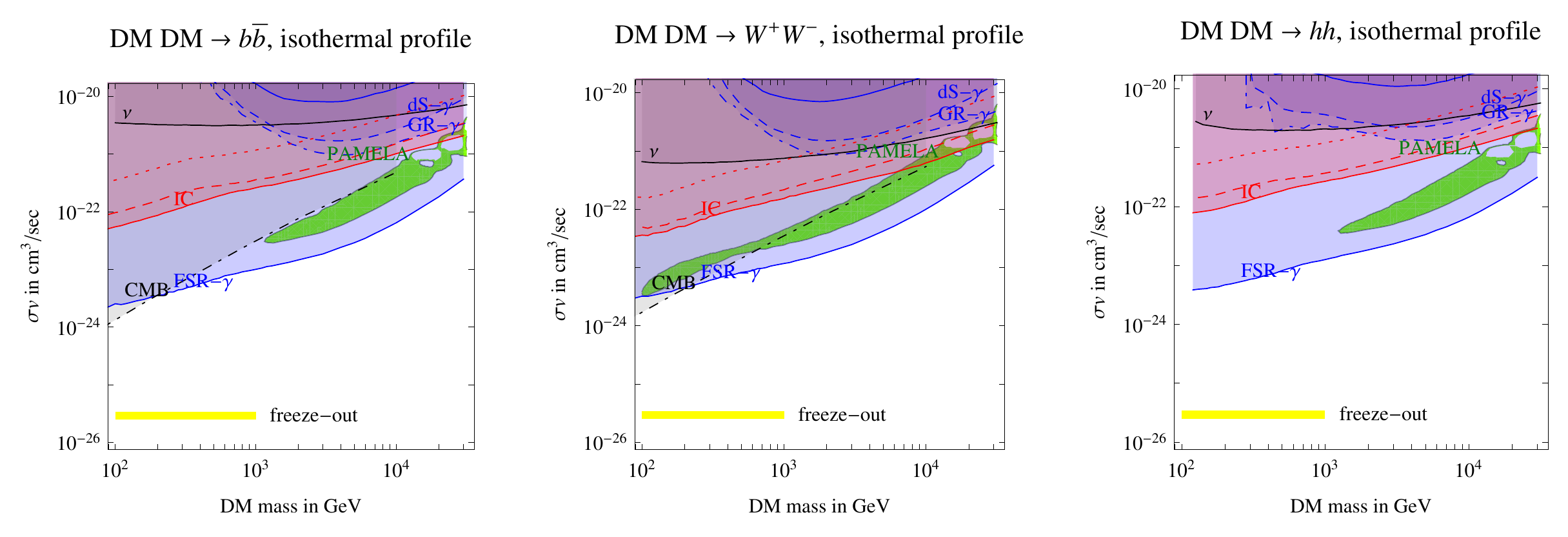}$$
$$\includegraphics[width=\textwidth]{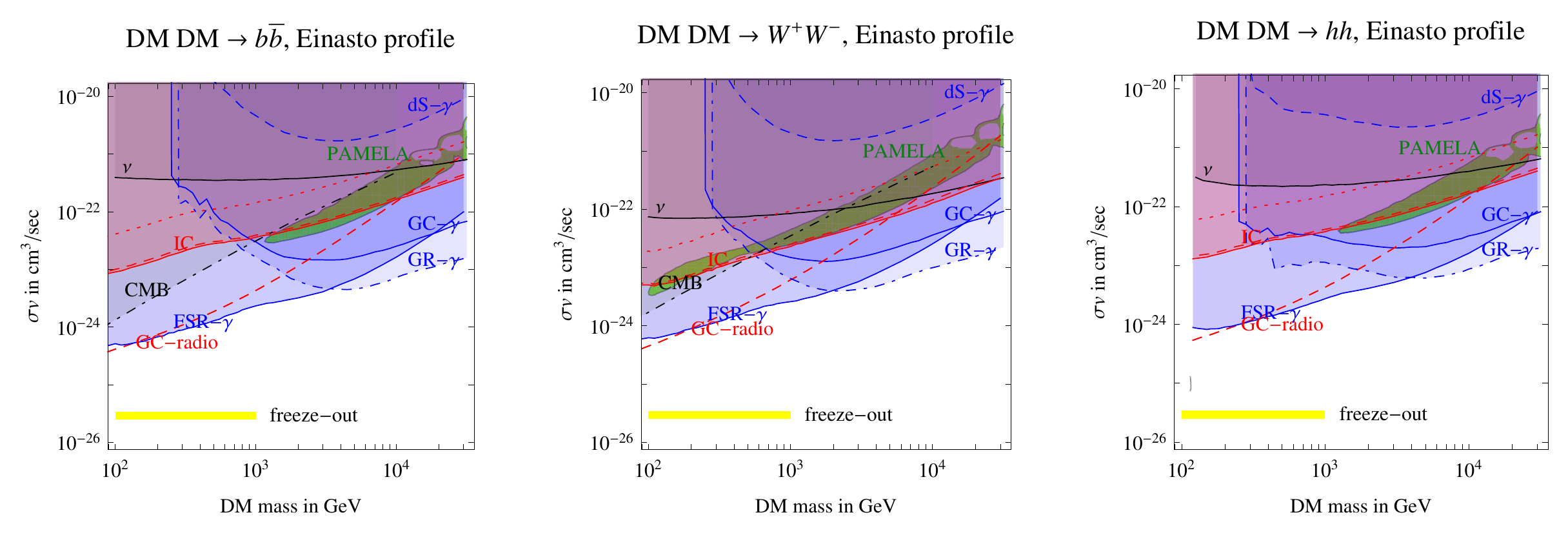}$$
\caption{\em {\bf Bounds on DM annihilations into non-leptonic channels}.
These channels can fit the PAMELA $e^+$ excess, but not the {\sc Fermi} $e^++e^-$ excess.
The {\sc Fermi} bounds on FSR$\gamma$ exclude non-leptonic DM interpretations of the PAMELA $e^+$ excess,
even for an isothermal DM profile. Non-leptonic branching ratios must be small.
}
\label{fig:annq}
\end{center}
\end{figure}

\section{DM annihilations} 
Assuming DM annihilations, fig.\fig{FSR} shows the {\sc Fermi} all-sky global bounds on FSR$\gamma$ 
as function of the DM mass $M$ and the DM cross section $\sigma v$.
Fig.\fig{IC} shows the corresponding bounds on IC$\gamma$, assuming $L=4\,{\rm kpc}$.

In both cases the left (right) panels holds for the isothermal (NFW) DM profile.
We see that the {\sc Fermi} bounds  only have a mild dependence on the DM profile:
although we do not know where DM is, {\sc Fermi} observed all the sky, so that it
is no longer possible to hide DM with an appropriate density profile.

The diffusion volume of $e^\pm$
is assumed to be a cylinder that extends away from the galactic plane up to $|z|<L$.
Fig.\fig{ICL} shows a typical example of how the IC bounds depend on $L$ and on the DM profile.
The main result is that if $L$ is as small as 1 kpc,
DM can produce a significant fraction of its $e^\pm$ outside of the diffusion volume,
that  negligibly contribute to the IC$\gamma$ signal.
Indeed they escape away, as the $e^\pm$ mean free path is one or two orders of magnitude longer than the Milky Way, and the probability
of entering into the diffusive halo is small.
We also checked the importance of the diffusion zone thickness with the more realistic model described in Sect.~\ref{astro}, in which diffusion exponentially dies off on a length scale $z_h$. 
Solving the diffusion equation in an ideally infinite volume, we find that the IC$\gamma$ flux is
closer to the large $L\approx 15$ kpc case even for $z_h=2\div 4$ kpc. 
Therefore any realistic bounds should be much closer to the ``max'' case than to the ones with very small $L$, especially since the energy dependence of the diffusion coefficient plays only a sub-leading role in the IC predictions. 

The possibility of a thin diffusion volume
will be relevant for our later discussion, so that it becomes important to settle this issue.
As far as we know, it is disfavored by various arguments: a) global fits of charged CR propagation models
favor $L\approx 4\,{\rm kpc}$ but values between 1 and 15 kpc are considered~\cite{minmedmax}.
b) abundances of CR with a life-time comparable to the diffusion time~\cite{MS}; 
c) more realistic boundary conditions as described above; and presumably d)
the fact that {\sc Fermi} observes $100\GeV$ $\gamma$ rays also away from the GC suggests that $L$ is not small.

\medskip

Fig.s\fig{ann} and\fig{annq} show again the {\sc Fermi} bounds at $3\sigma$
(the IC$\gamma$ bounds is plotted for a few values of the height of the diffusion volume, $L=1,2,4\,{\rm kpc}$), 
together with the regions favored by the $e^\pm$ excesses
and with various other $3\sigma$ bounds already considered in previous papers~\cite{BCST, HisanoNu,MPSV}:

\begin{itemize}
\item[-] The GC-$\gamma$ (blue continuous curves)
and GR-$\gamma$ (dot-dashed blue curves) bounds refer to the HESS observations~\cite{HessGC, HessGR}
of  the photon spectrum above $\approx 200\GeV$
(so that it constrains FSR$\gamma$ and heavier DM, rather than IC$\gamma$ and lighter DM)
in the `Galactic Center' region ($\sqrt{\ell^2 + b^2} < 0.1^\circ$)
and in the `Galactic Ridge' region ($|\ell| < 0.8^\circ$ and $|b| < 0.3^\circ$).
In these regions the DM density $\rho(r)$ is uncertain by orders of magnitude, such that
one gets  strong bounds assuming NFW-like
DM profiles and negligible bounds assuming isothermal-like profiles.

\item[-] The dS-$\gamma$ bound (dashed blue curves) refers to the HESS and VERITAS observations of various dwarf spheroidal galaxies~\cite{HessSgrDwarf, VERITAS,Essig}.

\item[-] The $\nu$ bounds (black curves)
refer to the SuperKamiokande (SK) observations of neutrino from regions around the Galactic Center~\cite{SK, HisanoNu,MPSV,Mandel}.

\item[-] The GC-radio bound (red dashed curves)
refers to radio observations of the Sgr A$^*$ black hole at the dynamical
center of the galaxy and depends on the extremely uncertain local DM density~\cite{Ullio, BCST}.

\item[-] The cosmological CMB bound (red dashed curves) refers to the contribution $\delta \tau$ to
the optical depth of CMB photons due to DM re-ionization of H and He\footnote{
We adopt the computation by Cirelli et al.~\cite{reionization} (not performed for all DM channels we consider),
who plotted the WMAP bound at $1\sigma$, $\delta\tau < 0.064$.
In our plots the region suggested by the PAMELA and {\sc Fermi} $e^\pm$ excesses is compatible with the CMB bound
because we plot the WMAP bound at $3\sigma$, $\delta\tau < 0.094$.
The CMB bound does not depend on the local DM density $\rho_\odot$, 
while all other curves actually constrain $\sigma v~\rho_\odot^2$.
Ref.~\cite{UllioCatena} claims that $\rho_\odot$ is larger than the
 $\rho_\odot=0.3\GeV/\cm^3$ assumed here and close to  $0.4\GeV/\cm^3$:
in such a case the CMB bound would be relatively less stringent by a factor of $1.8$.}.
\end{itemize}
In conclusion, DM interpretations of the $e^\pm$ excesses survive only if DM annihilates
into $2\mu$, $4\mu$ or $4e$ and if DM has a quasi-constant isotermal-like density profile.
The GC-$\gamma$ and GR-$\gamma$ bounds already disfavored solutions involving NFW or Einasto profiles~\cite{BCST,MPSV}, but this needed extrapolating these profiles down to small scales not probed by $N$-body simulations.
Now an unseen excess would be present at larger scales where $N$-body simulations are under control and favor these profiles.
Furthermore, channels involving $\tau$ are now disfavored even for an isothermal profile.
In view of the FSR-$\gamma$ FERMI bound, non-leptonic channels (fig.\fig{annq}) can similarly at most 
have a small sub-dominant branching ratio, 
so that solutions involving Minimal Dark Matter
or the supersymmetric wino~\cite{CKRS} are now firmly excluded.

The allowed solutions predict that a sizable fraction of the photons observed by {\sc Fermi} around 100 GeV
must be due to IC$\gamma$ from DM $e^\pm$.
The  {\sc Fermi} bound on IC$\gamma$
becomes weaker if the diffusive volume of our galaxy is thin, $L\approx 1\,{\rm kpc}$ (dotted red curves).

Another possible way of weakening the bound is assuming that a
fraction of the local DM density is stored in a dark disk component,
such that FSR$\gamma$ and (to a lesser extent) IC$\gamma$ are moved
towards the Galactic Plane, where the astrophysical $\gamma$ background is higher.
However, to relax the conclusions on the DM profile, one needs this fraction to be large, of order unity, especially if the dark disk has a thickness $z_d$ not much smaller than $r_\odot$.

On the other hand, the {\sc Fermi} bound can be made stronger subtracting from the $\gamma$ spectra the
hadrons misidentified as $\gamma$ and the identified astrophysical point-like sources.

 \begin{figure}[p]
\begin{center}
$$\includegraphics[width=\textwidth]{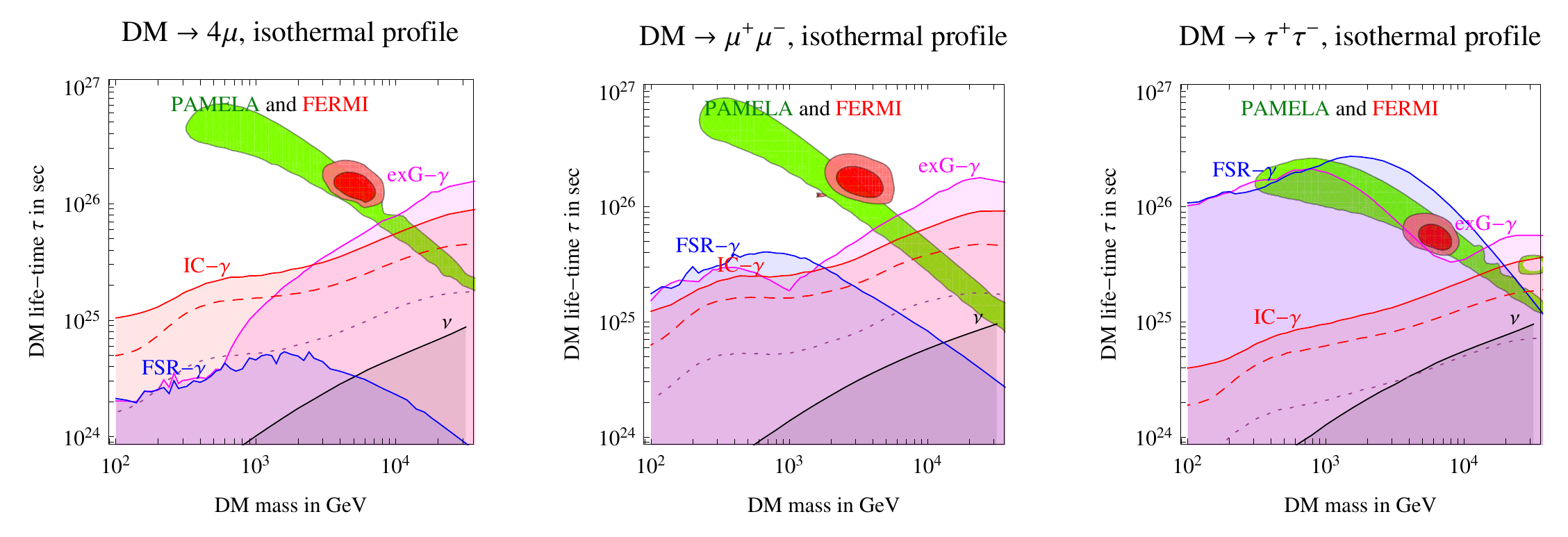}$$
$$\includegraphics[width=\textwidth]{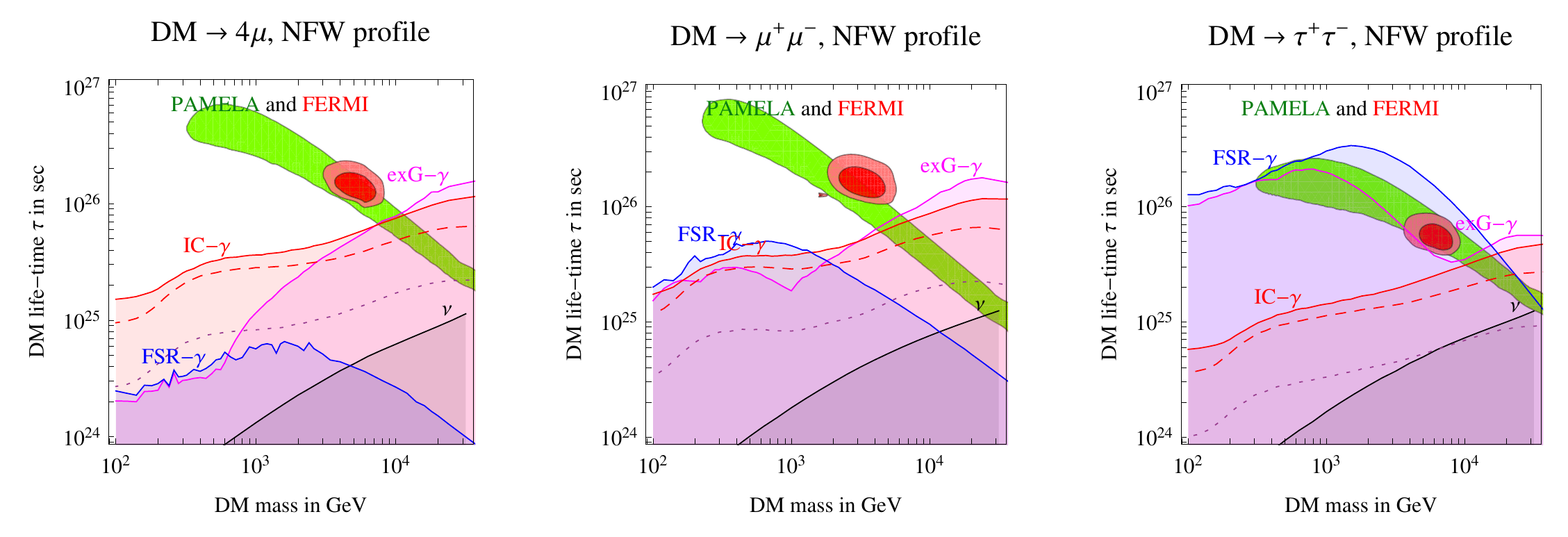}$$
$$\includegraphics[width=\textwidth]{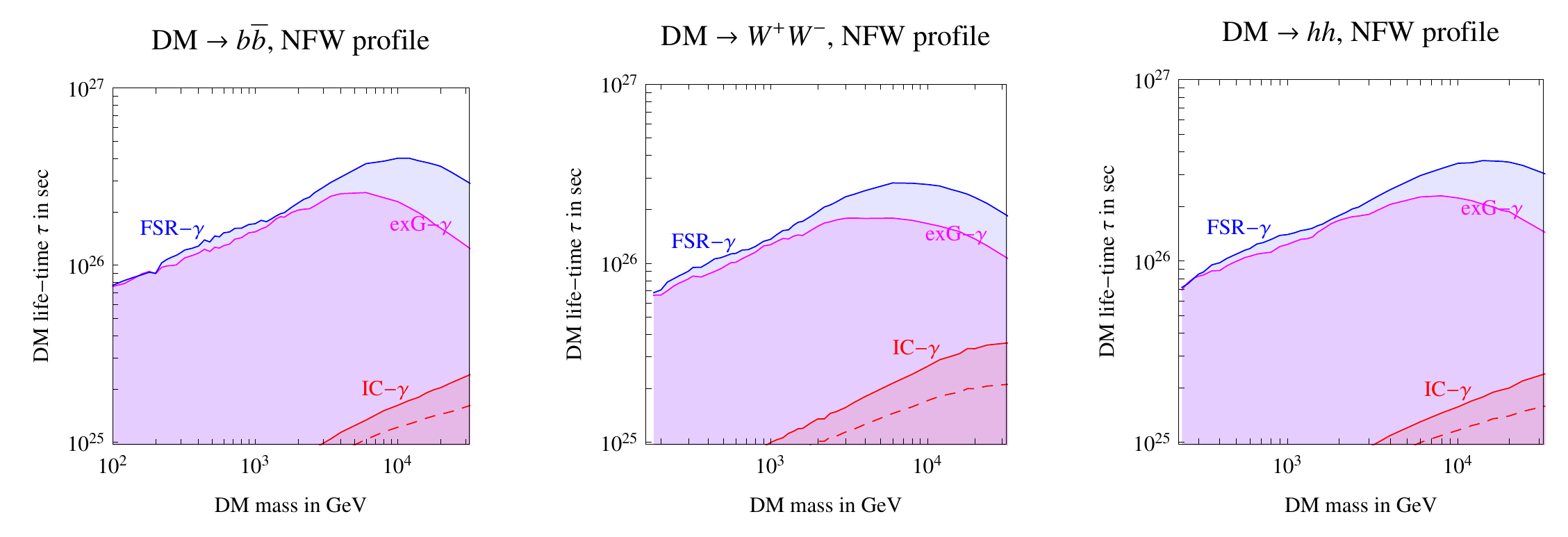}$$
\caption{\em {\bf Bounds on DM decays}. 
In the upper rows we consider the leptonic channels that can fit the $e^\pm$ excesses.
In the lower row we consider the `traditional' channels.
}
\label{fig:dec}
\end{center}
\end{figure}

\section{DM decays}
Interpretations of the $e^\pm$ excesses in term of DM decays (rather than annihilations)
attracted interest because they were not in tension with $\gamma$-ray observations~\cite{NSS}
(see also~\cite{decaying}).
Indeed, the space-time density of DM decays is $\rho/\tau M$ while the
space-time density of DM annihilations is $\sigma v (\rho/M)^2/2$.
Thereby, HESS observations of $\gamma$-rays from the  Galactic Center
gave significant constraints on DM annihilations if $r^3 \rho(r)^2$ is large for $r\to 0$.
On the contrary, even for a NFW profile, $r^3 \rho(r)$ remains small such that DM decays were not
significantly constrained.

\medskip

Fig.\fig{dec} shows, in the mass-lifetime plane, the new {\sc Fermi} FSR-$\gamma$ and IC-$\gamma$ bounds on DM decays,
together with previous bounds from SK neutrino observations
(for simplicity we do not plot the previous HESS $\gamma$ bounds, as they are now subdominant)
 and with the regions favored by  interpretations of the $e^\pm$ excesses in terms of DM decays.
We see that such interpretations are now constrained.
Also in the case of DM decays,  the remaining viable channels are $2\mu$, $4\mu$ or $4e$.
Channels like $\tau$'s, producing $\pi^0$ and other mesons decaying into photons,  are now disfavored.
Even in this case, a significant fraction of the photons observed by {\sc Fermi} around 100 GeV away from
the Galactic Center should be due to DM.

%


%

The bound denoted as  `exG-$\gamma$'  is obtained demanding that the cosmological $\gamma$ flux from
DM decays does not exceed the extra-galactic isotropic flux observed by {\sc Fermi}~\cite{FERMIdiffuse}.
This cosmological flux is expected to be comparable to the galactic flux:
\beq \frac{\Phi_{\rm cosmo}}{\Phi_{\rm galactic}} \sim \frac{\rho_{\rm cosmo} R_{\rm cosmo}}{\rho_\odot R_\odot}\sim 1\eeq
where $\rho_{\rm cosmo} = \Omega_{\rm DM} \rho_{\rm cr}\approx 1.3~10^{-6}\GeV/{\rm cm}^3$
and $R_{\rm cosmo}\sim 1/H_0\approx  13\,{\rm Gyr}$.\footnote{This bound was considered in~\cite{Ibarra,CiRun}
in the case of DM decays and in~\cite{Profumo} in the case of DM annihilations.
In such a case the larger suppression, $(\rho_{\rm cosmo} /\rho_\odot)^2$ is counteracted by DM clumping
in structures and galaxies, an effect which cannot be computed reliably.}
The isotropic cosmological $\gamma$ flux is
\beq\label{eq:cosmo} \frac{d\Phi_\gamma }{dE_\gamma} =
  \frac{c}{4\pi} \int_0^1 da \frac{e^{-\tau}a}{H(a)}\cdot \frac{dN_\gamma(E^{\rm in}_\gamma=E_\gamma/a)}{dV\,dt\,dE^{\rm in}_\gamma}\eeq
where the first term inside the integral generalizes the usual  line of sight integrand $ds$,
to the cosmological geometry described by the Hubble rate $H(a) = H_0 \sqrt{\Omega_\Lambda + \Omega_m/a^3}$,
as function of the scale factor $a$  of the universe, which gives the $E^{\rm in}_\gamma=E_\gamma/a$ redshift\footnote{Thereby the cosmological signal allows to probe DM decay models 
(not considered in this paper) that give hard photon emission 
only at $400\,\GeV-1\,\TeV$, as these photons, above the {\sc Fermi} reach, 
get partially red-shifted down to the {\sc Fermi} range.}.
The last term is the usual space-time density of $\gamma$ sources, equal to
$\Gamma (\rho_{\rm DM}(a)/M) dN_\gamma/dE^{\rm in}_\gamma$ in the case of DM decays.
We can neglect absorption of $\gamma$, as the optical depth is $\tau \ll 1$ below a few TeV.
Eq.\eq{cosmo} can be directly applied to the computation of FSR$\gamma$.

\smallskip

The flux of photons generated by cosmological Inverse Compton
scatterings of $e^\pm$  from DM annihilations
on Comic Microwave Background with energy density $u_\gamma(a)=\pi T^4/15$
and spectrum $dn_\gamma/dE = E^2/\pi^2/(e^{E/T}-1)$ at temperature $T=T_0/a$
can be written as:
\beq \frac{d\Phi_\gamma^{\rm IC}}{dE_\gamma} =\frac{9c m_e^4\Gamma  \rho_0}{32\pi H_0 M}
\int_0^1
\frac{da/a^2}{\sqrt{\Omega_\Lambda+\Omega_m/a^3}}   \int\!\! \!\!\int N_e(E_e^{\rm in})
\frac{1}{u_\gamma} \frac{dn_\gamma}{dE^{\rm in}_\gamma}
 \frac{dE_e^{\rm in}}{E_e^{\rm in 4}} \frac{dE^{\rm in}_\gamma}{E^{\rm in}_\gamma}f_{\rm IC}\ ,
\eeq
where $N_e(E)=  \int_{E}^{M}  dE'~{dN_{e}}/{dE'}  $ and
the function~\cite{ICth}
\beq
f_{\rm IC}=2q\ln q+(1+2q)(1-q)+\frac{1}{2}\frac{(\epsilon q)^2}{1+\epsilon q}(1-q)  \eeq
describes IC scattering
$\gamma(E_\gamma^{\rm in}) e(E_e^{\rm in})\to e\gamma(E^{\rm out}_\gamma = E_\gamma/a)$ at $E_e^{\rm in}\gg m_e$ 
in terms of the  dimensionless variables
\beq \epsilon=\frac{E^{\rm out}_\gamma}{E_e^{\rm in}},\qquad \Gamma=\frac{4E_\gamma^{\rm in} E_e^{\rm in}}{m_e^2},\qquad q= \frac{\epsilon}{\Gamma(1-\epsilon)}.\eeq
$E_{\gamma}^{\rm out}$ lies in the range $E_\gamma^{\rm in}/E_e^{\rm in} \le \epsilon\le \Gamma/(1+\Gamma)$.
The non-relativistic (Thompson) limit corresponds to $\Gamma\ll1$, so that $\epsilon\ll 1$ and $0\le q\le 1$.

\medskip

Fig.\fig{cosmosample}b shows the {\sc Fermi} isotropic $\gamma$ data
(according to the preliminary analysis in~\cite{FERMIdiffuse}), compared to the cosmological
flux generated by the best-fit DM $\to\tau^+\tau^-$ model.
In such a case the main constraint comes from FSR rather than from IC. 
Other DM decay models that can fit the $e^\pm$ excesses can have a much smaller
FSR$\gamma$, but have a very similar IC$\gamma$.
Thereby all such models predict an IC$\gamma$ at the level of the {\sc Fermi} observations.
Depending on how {\sc Fermi} extracted the isotropic component of their sky map,
it might be correct to include in it not only the extra-galactic DM$\gamma$ flux
but also the galactic DM$\gamma$ flux from the direction where it is minimal,
thereby strengthening the bound~\cite{CiRun}.

\section{Conclusions}
We presented robust model-independent bounds on DM annihilations and decays obtained demanding that the
computable part of the DM-induced $\gamma$ flux be below the observed flux,
as recently observed by {\sc Fermi} in the full sky.
Fig.\fig{mosaic} shows an example of bounds from the different regions of the sky we consider.
Our results in fig.s\fig{FSR} to\fig{annq} for DM annihilations and in fig.\fig{dec} for DM decays
are based on a full-sky global fit, and thereby only have a mild dependence on the DM density profile.

We show the bounds on Final State Radiation $\gamma$ separately from the bounds on
Inverse Compton $\gamma$, as the latter can be weakened if one allows for significant variations in astrophysics
with respect to the models we consider:
i) increasing the magnetic fields  until synchrotron energy losses dominate over IC; this is presumably 
allowed only  in the inner regions of the Galaxy, and would reduce our full-sky IC bounds by up to $1.5\div2$.
ii) IC bounds can be reduced by a factor of few if the $e^\pm$ diffusion zone is very thin ($L\sim1\,{\rm kpc}$) and terminates abruptly, see fig.\fig{ICL}.
iii) a Dark Disk component comprising $\sim50\%$ of the local DM density can also be invoked to weaken these bounds.

On the other hand, subtraction of astrophysical backgrounds
(such as identified point-like sources)
and of mis-identified hadrons, still present in the {\sc Fermi} data we fitted,
can only strengthen our bounds, presumably by a factor of few.

\medskip

Present data are enough to make progress on testing
DM interpretations of the $e^\pm$ excesses observed by PAMELA, {\sc Fermi}, ATIC.
At the light of the new {\sc Fermi} $\gamma$ bounds,
the remaining allowed DM interpretations involve DM annihilations or decays into
$\mu^+\mu^-$, $VV\to \mu^+\mu^-\mu^+\mu^-$ or $VV\to e^+e^-e^+e^-$ primary channels.
$\tau$'s in the final staes are now disfavored even in the DM decay case.
Moreover, for DM annihilation,  a quasi-constant isothermal-like density profile is needed.
This profile is not favored by DM simulations, that suggests that DM is concentrated 
around the Galactic Center.  In such a case, a viable interpretation may be obtained
e.g. assuming that DM annihilates into  intermediate particles
$V$ with a lifetime longer than a few kpc~\cite{Rothstein:2009pm}
or that DM decays.

Even in these cases, the expected DM signal is at the level of the observed flux 
so that it will be interesting to improve the sensitivity with forthcoming cleaner data and  more statistics. 
According to~\cite{Weiner}, present {\sc Fermi} $\gamma$ data already suggest the presence of a `{\sc Fermi} haze' excess
with an angular and energy spectrum compatible with the expected IC DM excess.


\subsubsection*{Acknowledgments}
We thank P. Meade, T. Volansky, M.\ Cirelli, P.\ Panci, P.D.\ Serpico for discussions. 
We verified that our bounds reduce to 
the ones in the recent paper~\cite{CiRun} if we neglect 
diffusion of $e^\pm$ in the Milky Way and the 
finiteness of the diffusion volume 
and restrict our {\sc Fermi} full-sky map to the three regions presented by {\sc Fermi} below 100 GeV.
The work of MP is supported in part by NSF grant PH0503584.

\appendix

\section{Extraction of the {\sc Fermi} data}

We used the {\sc Fermi} $\gamma$-ray data from~\cite{FERMIweb}. We selected events from the `diffuse' class, 
which have tighter cuts to reject the CR background. We considered the first 64 weeks of data (up to MET 280417908). After removing the Earth albedo, the data was binned in $0.5^\circ \times 0.5^\circ$ in galactic latitude and longitude and 
in 16 logarithmically spaced bins in energy. We computed the effective area and exposure times and used them to convert the binned events into the photon flux. We then combined the bins in the larger areas of Fig.~\ref{fig:mosaic}. We assigned systematic uncertainties according to~\cite{Porter:2009sg} for the energy bins below 100 GeV: 10\% at 100 MeV, 5\% at 0.5 GeV, progressively increasing to 20\% at 10 GeV and above. 

In this analysis we considered also events around and above 100 GeV: while these are still significantly contaminated by CR and the systematics of the publicly available tools are not completely studied, we feel that they are still usable for setting bounds the way we proceed in this paper. 
Only the possibility that the currently available software may overestimate the instrument effective area at high energies would render too stringent our bounds. Although this effect would be (partially) balanced by the reduction in the flux once the data is further cleaned from CR contamination, we conservatively decided to increase to 50\% the systematic uncertainty associated to all the data above 100 GeV.  

\bigskip
 
\footnotesize

\begin{multicols}{2}

\end{multicols}

\end{document}